\documentclass[showpacs,amsmath,nofootinbib,amssymb,twocolumn,a4paper,superscriptaddress,aps]{revtex4-1}

\usepackage{bm}
\usepackage{array}
\usepackage{color}
\usepackage[pdftex]{graphicx} 
\pdfoutput=1

\newcommand{\be}{\begin{equation}}
\newcommand{\ee}{\end{equation}}
\newcommand{\bea}{\begin{eqnarray}}
\newcommand{\eea}{\end{eqnarray}}

\newcommand{\Tb}{\ensuremath{\Tilde{\beta}}}
\newcommand{\iTb}{\ensuremath{\imath\Tilde{\beta}}}

\newcommand{\Hee}{\ensuremath{H_{ee}}}

\newcommand{\Hem}{\ensuremath{H_{e\mu}}}

\newcommand{\Hmm}{\ensuremath{H_{\mu\mu}}}

\newcommand{\Tk}[1]{\ensuremath{\Tilde{k}_{#1}}}

\newcommand{\dTk}[1]{\ensuremath{\delta\Tilde{k}_{#1}}}
\newcommand{\Bk}[1]{\ensuremath{\breve{k}_{#1}}}

\newcommand{\dBk}[1]{\ensuremath{\delta\breve{k}_{#1}}}
\newcommand{\dTkd}[1]{\ensuremath{\delta\dot{\Tilde{k}}_{#1}}}
\newcommand{\dBkd}[1]{\ensuremath{\delta\dot{\breve{k}}_{#1}}}

\newcommand{\Tt}[1]{\ensuremath{\Tilde{\theta}_{#1}}}

\newcommand{\TU}{\ensuremath{\Tilde{U}}}
\newcommand{\TUd}{\ensuremath{\Tilde{U}^{\dagger}}}

\newcommand{\Pvec}{\ensuremath{\mathbf{P}}}
\newcommand{\Bvec}{\ensuremath{\mathbf{B}}}
\begin{document}
\topmargin0.1in
\bibliographystyle{apsrev}
\title{The neutrino spectral split in core-collapse supernovae: a magnetic resonance phenomenon}

\author{S\'ebastien Galais}
\email{galais@ipno.in2p3.fr}
\affiliation{Institut de Physique Nucl\'{e}aire Orsay, F-91406 Orsay cedex, France}

\author{Cristina Volpe}
\email{volpe@ipno.in2p3.fr}
\affiliation{Institut de Physique Nucl\'{e}aire Orsay, F-91406 Orsay cedex, France}

\date{\today}

\pacs{14.60.Pq,,97.60.Bw}

\begin{abstract}
A variety of neutrino flavour conversion phenomena occur in core-collapse supernova, due to the large neutrino density close to the neutrinosphere, and the importance of the neutrino-neutrino interaction. Three different regimes have been  identified so far, usually called the synchronization, the bipolar oscillations and the spectral split. Using the formalism of polarization vectors, within two-flavours, we focus on  the spectral split phenomenon and we show for the first time that the physical mechanism underlying the neutrino spectral split is a magnetic resonance phenomenon. In particular, we show that the precession frequencies fulfill the magnetic resonance conditions. Our numerical calculations show that the neutrino energies and the location at which the resonance takes place in the supernova coincide well with the neutrino energies at which a spectral swap occurs. The corresponding adiabaticity parameters present spikes at the resonance location.
\end{abstract}

\maketitle

\section{Introduction} \label{sec:intro}
\noindent
Intense theoretical activity has unravelled the complexity of neutrino flavour conversion in media. The neutrino propagation in a star like our sun is at present well understood as a resonant flavour conversion phenomenon due to the coupling of neutrinos with ordinary matter, the Mikheev-Smirnov-Wolfenstein (MSW) effect \cite{Wolfenstein1977,M&S1986}. This confirmation has in particular been brought by the 
SNO \cite{Ahmad:2002jz}, KamLAND \cite{Eguchi:2002dm} 
and recently by the Borexino \cite{Bellini:2008mr} 
experiments.
Neutrino oscillations in massive stars turn out to be a non-linear phenomenon because of the importance of the neutrino-neutrino interaction, as first pointed out in \cite{Pantaleon1992eq,Samuel:1993uw} and also later on \cite{Kostelecky:1993ys,Kostelecky:1994dt,Balantekin:2004ug}. 
In particular, collective flavour conversion phenomena have been shown to emerge \cite{Duan:2005cp}, the understanding of which has already required a large theoretical effort (see e.g. 
\cite{Samuel:1993uw,Balantekin:2004ug,Duan:2005cp,Dasgupta:2007ws,
Hannestad:2006nj,Balantekin:2006tg,Duan:2006an,Raffelt:2007xt,Raffelt:2007cb,
Gava:2008rp,Fogli:2009rd,Gava:2009pj,Dasgupta:2009mg,Duan:2010bf,Galais:2011jh}) and \cite{Duan:2010bg,Duan:2009cd} for a review). 
Besides, in such explosive environments the presence of shock waves 
\cite{Schirato:2002tg,Takahashi:2002yj,Fogli:2003dw,Tomas:2004gr,Choubey:2006aq,Kneller:2007kg,Gava:2009pj} 
and of turbulence \cite{Loreti:1995ae,Fogli:2006xy,Friedland:2006ta,Kneller:2010sc} 
also induces new flavour conversion effects. 
The inclusion of all these features has significantly increased the complexity in the computation of neutrino propagation in such media. Fully understanding the flavour conversion phenomena and their interplay with key unknown neutrino properties, such as the neutrino mass hierarchy, the third unknown mixing angle and leptonic CP violation 
\cite{Balantekin:2007es,Gava:2008rp} is one of the major goals in this domain. While important progress has been achieved, several aspects still need further clarification.

\begin{figure}[h]
	\centering
		\includegraphics[width=90mm]{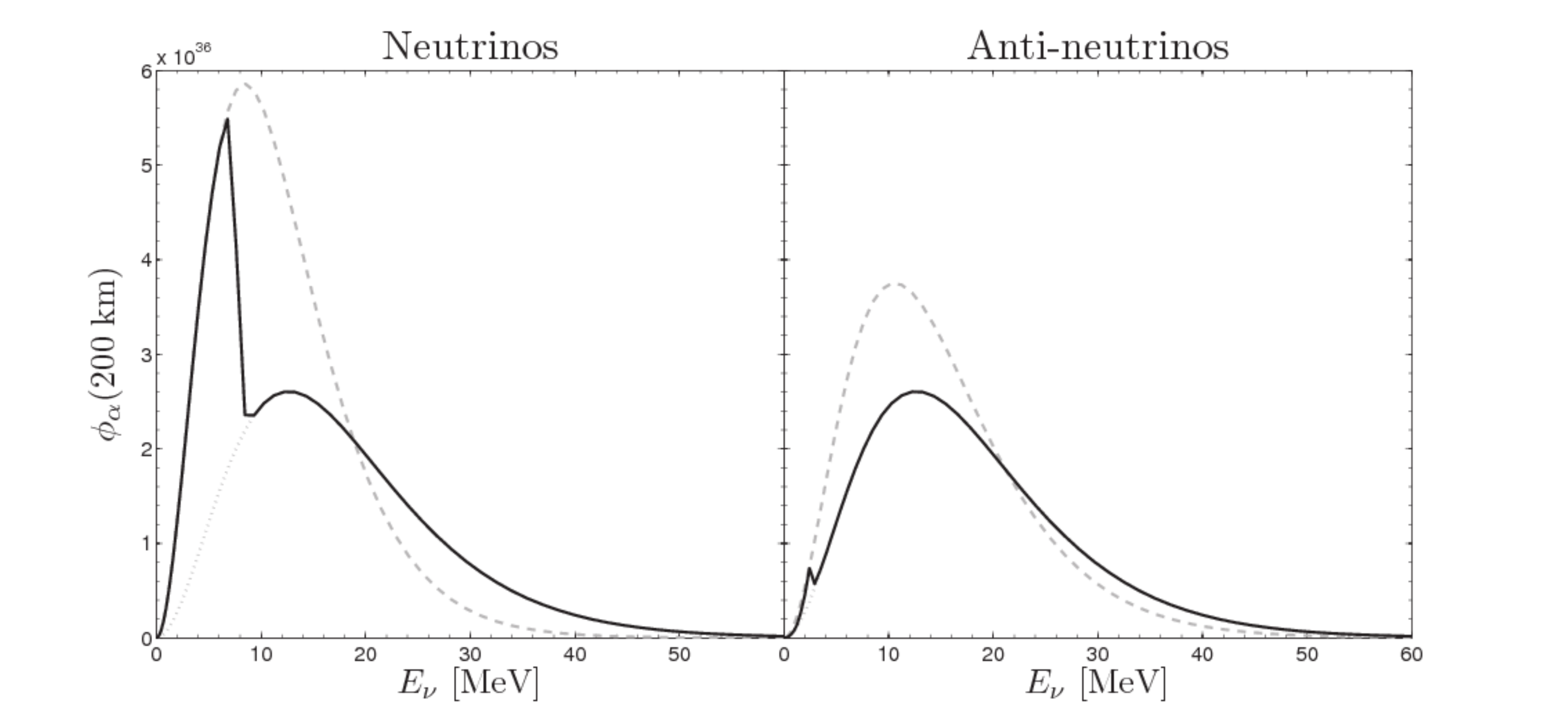}
	\caption{Neutrino fluxes as a function of neutrino energy at 200 km near the neutrinosphere, in the case of inverted hierarchy and within two neutrino flavours. The spectral split due to the neutrino-neutrino interaction is apparent as a swap of the electron neutrino with the $\nu_x$ neutrino spectra. The results correspond to neutrinos (left figure), where the split energy is at 7.6 MeV, and anti-neutrinos (right figure), with a split energy at about 2.4 MeV.}
	\label{fig:flux}
\end{figure}

Currently, three collective flavour conversion regimes induced by the interaction of neutrinos with themselves are identified, usually called the synchronization regime, the region of bipolar oscillations and the spectral split phenomenon. The first occurs very close to the neutrinosphere : any flavour conversion is frozen due to the neutrino-neutrino interaction \cite{Duan:2005cp}. 
Then the vacuum term triggers an instability  
\cite{Duan:2005cp,Hannestad:2006nj,Duan:2010bf,Galais:2011jh} (depending on the neutrino luminosities and the hierarchy) 
that produces flavour oscillations. 
It appears to be associated with a rapid growth of the derivative of the matter phase introduced by the neutrino-neutrino interaction \cite{Galais:2011jh}.
The onset of such an instability is different in calculations based on the single-angle approximation versus a multi-angle treatment \cite{Duan:2010bf}. 
Bipolar oscillations have been shown to correspond to electron anti-neutrino and neutrino pairs converting to pairs of neutrinos and anti-neutrinos of other flavours \cite{Hannestad:2006nj}. 
Finally, as first indentified in \cite{Duan:2005cp} the neutrinos undergo a complete flavour conversion depending on their energy. This implies the swap of the electron neutrino spectra with the ones of muon and tau neutrinos, below or above a critical value called the split energy. The spectral split has been further studied in 
\cite{Raffelt:2007xt,Raffelt:2007cb} (Figure \ref{fig:flux}) and   
Refs. \cite{Dasgupta:2009mg,Fogli:2009rd} 
have shown that such a phenomenon is even more complex since multiple spectral splits show up, depending on the neutrino flux ratios at the neutrinosphere and on the core-collapse supernova explosion phase. While several aspects have now been identified, a comprehensive understanding of the spectral split phenomenon is still missing. 

Extensively used in the literature, the formalism of neutrino polarization vectors constitutes a good tool to gain a better understanding of neutrino flavour conversion phenomena. A long time ago it was pointed out that neutrino oscillation 
in vacuum could be pictured as the precession of a neutrino flavour polarization vector around an effective magnetic field, depending on the neutrino masses and mixings \cite{Stodolsky:1986dx}. 
Later on \cite{Kim:1987bv,Kim:1987ss} 
the case of the resonant (adiabatic and non-adiabatic) flavour conversion in matter has been discussed, in connection with the MSW effect and the solar neutrino deficit problem. 
Recently such a formalism has been used in the presence of the neutrino-neutrino interaction Hamiltonian in supernovae first in \cite{Duan:2005cp} and subsequently in \cite{Hannestad:2006nj,Raffelt:2007xt,Raffelt:2007cb,Dasgupta:2007ws,Duan:2007fw}. 

Using such a formalism in Ref.\cite{Pastor:2001iu} 
the synchronization regime has been investigated in the context of the Early Universe. 
The bipolar oscillation regime has been put in relation with a flavour pendulum in \cite{Duan:2005cp,Duan:2007mv}
and with a gyroscopic pendulum in \cite{Hannestad:2006nj}. 
In \cite{Duan:2007mv,Duan:2007fw} the authors make the hypothesis that at the end of the synchronization and of 
the bipolar regimes, the neutrino evolution follows a collective precession mode until neutrino densities are low 
and at the late stage of this precession solution the stepwise swapping occurs.
In \cite{Raffelt:2007cb}, an explicit adiabatic solution of the two flavour neutrino evolution equations with the 
neutrino-neutrino interaction is built in the comoving frame. The existence of such a frame had already been pointed 
out in the  early work \cite{Duan:2005cp}. 
It is shown that analytical solutions exist and present a behaviour like the one of the neutrino spectral split. 
 Approximate conserved quantities, in particular the total neutrino lepton number for the two-flavour case, have been 
identified and have been shown to explain the split energy observed in the neutrino fluxes \cite{Raffelt:2007cb}. 
This idea has been further investigated in \cite{Raffelt:2007xt}.
Besides the synchronization and the pure precession modes already mentioned, a self-induced parametric resonance
mode is discussed in \cite{Raffelt:2008hr}.
Many of these studies have brought very valuable insights into the collective effects engendered by the neutrino 
self-interaction and have given a vision of what actually is seen in the full numerical calculations. 
Note that, in order to allow for these useful
analytical solutions, approximations are often made : (i) neglecting the matter term; (ii) using a constant coefficient 
for the neutrino self-interaction term or with a simplified time dependence; or (iii) employing a single neutrino or 
anti-neutrino energy or neutrino box spectra.
An application of the polarization vector formalism to the three-flavour case, requiring the $SU(3)$ basis, is performed in \cite{Dasgupta:2007ws} 
and it has been shown 
that two conserved quantities also appear in this case \cite{Duan:2007fw,Duan:2008za}.
Finally, adiabaticity of the neutrino evolution has been particularly discussed in \cite{Duan:2007fw,Raffelt:2007cb,Duan:2008za}.
  
In this work we approach the problem of the neutrino spectral split using the matter basis and the polarization vector formalism within two neutrino flavours. We give the explicit analytical expressions for the
polarization vectors and the effective magnetic field, that in our case includes both the neutrino-neutrino interaction, the vacuum contribution and also the matter term. In the matter basis we follow the full neutrino evolution in the region where the neutrino-neutrino interaction effects are dominating, and focus, in particular, on the spectral split phenomenon. We show, for the first time, that a magnetic resonance mechanism is the underlying process governing the flipping of the polarization vectors and therefore the spectral split.  
We point out that the flipping -  giving rise to a spectral swapping - occurs at the neutrino energies for which the precession frequencies meet the magnetic resonance condition. We show the behaviour of the associated generalised adiabaticity parameters.   

The paper is structured as follows. Section II recalls the polarization vector formalism for two-neutrino flavours in connection with the neutrino Hamiltonian describing their propagation in a core-collapse supernova. Section III
gives the equations to follow the evolution of the effective magnetic field and of the polarization vector.  Section IV establishes the link between the spectral split phenomenon and a magnetic resonance. It presents the numerical results for the neutrino probabilities and adiabaticity parameters in the region where the neutrino-neutrino interaction effects are important, and shows that the resonance condition is indeed satisfied when the spectral split occurs. Section V is a conclusion. In the Appendix, expressions to calculate the second derivative of the neutrino-neutrino interaction Hamiltonian in the multi-angle treatment, are provided.

\section{The polarization vector formalism and neutrino evolution in a supernova in the matter basis} 
\noindent
Let us start by reminding the well known fact that the quantum evolution of a system described by an Hermitian $2 \times 2$~matrix 
\begin{equation}\label{e:1}
H =\left( \begin{array}{cc}
		H_{11} & H_{12}  	\\
		H_{21}  & H_{22}   \end{array} \right) \,
\end{equation}
with $H_{21}= H_{12}^*$, can be associated with an effective magnetic field   \cite{Cohen}
\begin{equation}\label{e:2}
\bf{B} =
\left( \begin{array}{c}
	 B_x 	\\		
     B_y 	\\
        B_z          
 \end{array} \right) = 
\left( \begin{array}{c}
		2 \mathcal{R}(H_{12})  	\\		
        - 2  \mathcal{I}(H_{12})  	\\
         (H_{11}-H_{22})              \end{array} \right)
\end{equation}
with $\mathcal{R}$ and $\mathcal{I}$ indicating the real and imaginary contributions of the off-diagonal matrix element.
The evolution of the system can be seen as the precession of an effective spin 
around $\Bvec$. 
Indeed, if we define the polarization vector $\bf{P}$ 
\begin{equation}\label{e:4}
\bf{P}=  \langle \psi | \boldsymbol{\sigma} | \psi  \rangle .
\end{equation}
with $\boldsymbol{\sigma}$ the Pauli matrices,
the evolution equation\footnote{Note that in the manuscript we will use $x$ to denote time.} can be written as
\begin{equation}\label{e:precession}
{{d \bf{P}} \over {dx}} = \Bvec \times \bf{P}
\end{equation}
with precession frequency $ \omega = |\Bvec|$. 

Let us now consider the specific case of the neutrino evolution in a core-collapse supernova. The corresponding Hamiltonian in the flavour basis is :
\begin{equation}\label{e:5}
H^{(f)} = U\,K\,U^{\dag}+ H_{m}^{(f)}+ H^{(f)}_{\nu\nu}
\end{equation} 
where the non-diagonal vacuum contribution is $U K U^{\dagger}$ with $K = diag (k_1,k_2)$, $U$ is the Maki-Nakagawa-Sakata-Pontecorvo (MNSP) unitary matrix, relating the flavour to the mass basis  \cite{Nakamura:2010zzi}. The other two contributions correspond to
the matter term\footnote{Note that the neutral current contribution is taken out since it only contributes with a term proportional to the unity matrix and is therefore irrelevant for neutrino oscillations.} $H_m^{(f)} = \sqrt{2} G_F \rho_e$, coming from charged-current interactions of neutrinos with the electrons in the medium, and the neutrino-neutrino interaction term $H^{(f)}_{\nu\nu}$  \cite{Duan:2006an}. 
We will follow the neutrino evolution in the matter basis\footnote{From now on, the symbol $\tilde{O}$ will always indicate that 
the quantity (in this case $O$) is calculated in the matter basis.}, as done in Ref.\cite{Galais:2011jh}.
In such a basis the Hamiltonian is :
\begin{equation}\label{e:6}
\tilde{K} = \tilde{U}^{\dagger} H^{(f)} \tilde{U} = diag(\tilde{k}_1,\tilde{k}_2)
\end{equation} 
with the two-neutrino flavour Hamiltonian
\begin{equation}\label{e:6bis}
H^{(f)}  = \left( \begin{array}{cc}
		H_{ee} & H_{e \mu}  	\\
		H_{\mu e}  & H_{\mu\mu}   \end{array} \right) \,
\end{equation} 
and $\tilde{U}$ the MNSP matrix depending on the matter angles and phases,
$(\tilde{k}_1,\tilde{k}_2)$ the matter eigenvalues. 
The evolution equation reads
\begin{equation}\label{e:evm}
\imath {{d \tilde{\psi}} \over dx} = \tilde{H} \tilde{\psi}
\end{equation}
with $\tilde{\psi}=(\nu_1,\nu_2)$ the matter eigenstates and the matter Hamiltonian involving the derivatives of the $\tilde{U}$ matrix \cite{Galais:2011jh}
\begin{equation}\label{e:Hm}
\tilde{H} = \tilde{K} - \imath   \tilde{U}^{\dagger} {d \tilde{U} \over dx} .
\end{equation}
More explicitly one gets :
\begin{equation}\label{e:Hm2}
\tilde{H} = \left( \begin{array}{cc}
		\Tk{1}+Q_{1}  & \imath\,\frac{(\dTk{12}+\delta Q_{12})}{2\,\pi}\,\Gamma  	\\
		-\imath\,\frac{(\dTk{12}+ \delta Q_{12})}{2\,\pi}\,\Gamma^{\ast} & \Tk{2}+Q_{2}   \end{array} \right) \,
\end{equation}
with $\dTk{12}= \tilde{k}_1 -  \tilde{k}_2$
the diagonal contributions coming from the $\tilde{U}$ derivative 
\begin{equation}\label{e:Qi}
Q_{i}= - \imath \left(\TU^{\dag}{\frac{d\TU}{dx}}\right)_{ii}
\end{equation}
with $i=1,2$, and we define
\begin{equation}\label{e:15}
\Bk{i} = \Tk{i} -\imath\left(\TUd\dfrac{d\TU}{dx}\right)_{ii}.
\end{equation}
The off-diagonal terms are given by the generalized adiabaticity parameters \cite{Galais:2011jh} 
\begin{eqnarray}\label{e:Gammaij}
\Gamma & = & -\frac{2\,\pi e^{\imath \tilde{\alpha}_{12}}}{(\dTk{12} + \delta Q_{12})}\left(\TU^{\dag}{\frac{d\TU}{dx}}\right)_{12} \\ \nonumber
& = & \frac{2\,\pi e^{\imath \tilde{\alpha}_{12}}}{\dTk{12}(\dTk{12} + \delta Q_{12})} \left(\TU^{\dag}{\frac{dH}{dx}} \TU \right)_{12}.
\end{eqnarray}
We take the most general expression for $\TU$ :
\begin{equation}\label{e:Um}
\TU = 
\left(\begin{array}{ll} 1 & 0  \\ 0 & e^{\iTb} \end{array}\right)
\left(\begin{array}{ll} \cos\Tt{} & \sin\Tt{}  \\ -\sin\Tt{} & \cos\Tt{} \end{array}\right)
\left(\begin{array}{ll} e^{- \imath \tilde{\alpha}_1} & 0 \\ 0 &  e^{- \imath \tilde{\alpha}_2}\end{array}\right)
\end{equation}
where $\Tb$ and $\tilde{\alpha}_1,\tilde{\alpha}_2$ are the Dirac and Majorana phases in matter respectively. 
The neutrino evolution equation becomes explicitly \cite{Galais:2011jh} :
\begin{equation}\label{e:Hm2f}
\imath {d \over dx}\left( \begin{array}{c}
	\nu_1 	\\		
	\nu_2	  \end{array}  \right) 
=    \left( \begin{array}{cc}
		\tilde{k}_1 + \dot{\tilde{\beta}}\rm{sin}^2 \tilde{\theta}  & - \dot{\tilde{\beta}}
\frac{\rm{sin}2 \tilde{\theta}}{2}- i \dot{\tilde{\theta}}	\\
		 - \dot{\tilde{\beta}}\frac{\rm{sin}2 \tilde{\theta}}{2}+ i \dot{\tilde{\theta}}  & \tilde{k}_2 + \dot{\tilde{\beta}}\rm{cos}^2 \tilde{\theta}  \end{array} \right) \left( \begin{array}{c}
	\nu_1 	\\		
	\nu_2	  \end{array}  \right) 
\end{equation}
where we use the dot to indicate $d/dx$. The Majorana phases $(\tilde{\alpha}_1,\tilde{\alpha}_2)$ are set to zero without loss of generality since they do not influence the neutrino flavour conversion.

We now apply the polarization vector formalism to Eq.(\ref{e:Hm2}). 
Following Eqs.(\ref{e:1}-\ref{e:2}), one gets the effective magnetic field corresponding to the matter Hamiltonian 
\begin{equation}\label{e:B}
\tilde{\bf{B}} = 
\left( \begin{array}{c}
- {{\Bk{12}}\over{\pi}}\mathcal{I}(\Gamma)		\\		
 - {{\Bk{12}}\over{\pi}}\mathcal{R}(\Gamma)	    	\\
        \Bk{12}     
 \end{array} \right) =
\left( \begin{array}{c}
	 -   \dot{\tilde{\beta}}\rm{sin}2 \tilde{\theta}	\\		
    2 	\dot{\tilde{\theta}} \\
  \delta \tilde{k}_{12} - \dot{\tilde{\beta}}\rm{cos}2 \tilde{\theta}
 \end{array} \right)
\end{equation}

The magnetic field x- and y-components only depend upon the Hamiltonian off-diagonal element, while the z-component is the difference of the diagonal ones. 
From Eq.(\ref{e:4}) one can construct a neutrino polarization vector in the matter basis 
\begin{equation}\label{e:P}
\tilde{\bf{P}} = 
\left( \begin{array}{c}
	 2 \mathcal{R}(\nu_1\nu_2^*)	\\		
   - 2 \mathcal{I}(\nu_1\nu_2^*)	\\
 |\nu_1|^2 -|\nu_2|^2     
 \end{array} \right) 
\end{equation}
whose z-component tells us the $\nu_1$ or $\nu_2$ content of the neutrino wavefunction. The x- and y-components are as usual non-zero in presence of mixings.

Now solving Eq.(\ref{e:Hm2f}) is equivalent to determining 
the evolution of the polarization vector Eq.(\ref{e:P}) under the action of the effective magnetic field Eq.(\ref{e:B}), i.e. Eq.(\ref{e:precession}). In this work,
we retain all contributions to the Hamiltonian including the matter term. In order to follow the $\bf{P}$ evolution we need to calculate the matter angle derivative 
\begin{equation}\label{e:derangle}
\dot{\tilde{\theta}}=\frac{\left( \dot{H}_{ee}-\dot{H}_{\mu\mu}\right)\,\dTk{12} - (\Hee-\Hmm)\,\delta\dot{\tilde{k}}_{12}}{4|\Hem|\dTk{12}}
\end{equation} 
and the phase derivative  :
\begin{equation}\label{e:derphase}
 \dot{\tilde{\beta}}=\dfrac{1}{|\Hem|^2}\left[\mathcal{I}(\Hem)\mathcal{R}(\dot{H}_{e\mu})- \mathcal{R}(\Hem)\mathcal{I}(\dot{H}_{e\mu})\right]
\end{equation} 
Eqs.(\ref{e:derangle}-\ref{e:derphase}) involve the derivatives of the diagonal $H_{ee}$, $H_{\mu\mu}$ and off-diagonal $H_{e\mu}$ Hamiltonian matrix elements in the flavour basis Eq.(\ref{e:6bis})  (see \cite{Galais:2011jh} for details).  Note that, if matter only is included, only the derivative of the matter angle is relevant, while the presence of complex off-diagonal contributions in the Hamiltonian in the flavour basis introduces also the $\tilde{\beta}$ phase and its derivative.

\section{The evolution of $\Bvec$ and $\Pvec$} 

\noindent
In order to establish a connection between the spectral split phenomenon and the magnetic resonance phenomenon let us now define relevant quantities.  
First of all
the total precession frequency of the polarization vector Eq.(\ref{e:P}) around the effective magnetic field Eq.(\ref{e:B}) is
\begin{equation}\label{e:omegaP}
\tilde{\omega}_{\bf P} = {\bf |  \tilde{\Bvec}  |},
\end{equation} 
while the angle defining the evolution of $\tilde{\Bvec}$ with respect to the (XOY) plane is defined as 
\begin{equation}\label{e:taneta}
 \tan \eta = \dfrac{\tilde{\Bvec}_z }{ \left|  \tilde{\Bvec}_\perp  \right|} = \dfrac{ \dBk{12} }{\left| 2 \tilde{H}_{12}    \right|} .
\end{equation}
In order to follow the evolution of the system it
is interesting to define the rate of change of the $\tilde{\Bvec}$ direction with respect to such a plane, namely 
\begin{eqnarray}\label{e:omegaeta}
 \tilde{\omega}_\eta = \dot{\eta} &= &2 f_{\eta} \left[\delta\dot{\breve{k}}_{12} | \tilde{H}_{12}| - 2{\dBk{12}} \left[ \mathcal{R}({\tilde{H}_{12}}) \mathcal{R}({\dot{\tilde{H}}_{12}}) \right.\right.   \nonumber \\
&&\left.\left.  + \mathcal{I}({\tilde{H}_{12}}) \mathcal{I}({\dot{\tilde{H}}_{12}})  \right] | \tilde{H}_{12}|^{-1}\right]
\label{eq:omega_eta}
\end{eqnarray}
with the normalization factor 
$f_{\eta}=[ 4 | \tilde{H}_{12}|^2 + \dBk{12}^2 ]^{-1}$.
One sees that to follow $\tilde{\omega}_\eta$ both the derivative of $\dBk{12}$ and 
of $\Gamma$ are needed. For the former, one gets 
\begin{equation}\label{e:23}
\delta \dot{\breve{k}}_{12}  = \dot{\tilde{\Bvec}}_z = 
\delta \dot{\tilde{k}}_{12} - \ddot{\tilde{\beta}} \cos2\Tt{} + 2\dot{\Tb}\dot{\Tt{}} \sin2\Tt{} 
\end{equation}
with
\begin{equation}\label{e:24}
 \delta \dot{\tilde {k}}_{12} = \left( \TUd\dot{H}^{(f)}\TU\right)_{11} -  \left( \TUd\dot{H}^{(f)}\TU\right)_{22}
\end{equation}
while one gets for the second derivative of $\tilde{\beta}$
\begin{eqnarray}\label{e:25}
 \ddot{\Tb} &=&  f_{\beta} \left[ \mathcal{I}({\Hem})\,\mathcal{R}(\ddot{H}_{e\mu})
 -\mathcal{R}({\Hem})\,\mathcal{I}(\ddot{H}_{e\mu})\right]  \\ 
&&-2\,f_{\beta}\,\dot{\tilde{\beta}}\,\left[ \mathcal{R}(\Hem)\,\mathcal{R}(\dot{H}_{e\mu})+\mathcal{I}(\Hem)\,
\mathcal{I}(\dot{H}_{e\mu})\right] \nonumber
\end{eqnarray}
with $f_{\beta}=|\Hem|^{-2}$. 
The other important quantity to determine for Eq.(\ref{e:omegaeta}) is
\begin{eqnarray}\label{e:derGamma}
\dTk{12}\dBk{12}\,\dot{\Gamma} &=& -
\left(\dTkd{12}\dBk{12}+\dTk{12}\dBkd{12} \right) \Gamma \nonumber\\
&&+ 2\pi \left( \TUd\ddot{H}^{(f)}\TU + \left[ \TU^{\dagger}
\dot{H}^{(f)}\TU,\TUd\dot{\tilde{U}} \right] \right)_{12}
\end{eqnarray}
which depends upon the second derivative of $H^{(f)}$ Eq.(\ref{e:5}).
Such a derivative involves two contributions, 
since the vacuum term is constant
\begin{eqnarray}\label{e:d2H}
 \ddot{H}^{(f)} & = & \ddot{H}^{(f)}_{m} + \ddot{H}^{(f)}_{\nu\nu} 
\end{eqnarray}
The first term depends on the explicit matter density profile used, that we take as a power-law. Its contribution to Eq.(\ref{e:d2H}) is straightforward. As far as the neutrino-neutrino interaction term is concerned, the corresponding second derivative depends on the use of a single-angle approximation versus a multi-angle treatment. 
Here we give its expression in the former case since it is the approximation we employ in the following. The relations valid in the multi-angle case are given in the Appendix.
The neutrino-neutrino interaction Hamiltonian in the single-angle approximation\footnote{Note that in our calculations the neutrinos are emitted with the same angle taken to be $0^{\circ}$ with respect to the neutrinosphere as done e.g. in Ref. \cite{Duan:2006an}.} is given by
\begin{equation}\label{e:24b}
H^{(f)}_{\nu\nu}=F(x)G(\rho)
\end{equation} 
where $\rho$ indicates the density matrix for two neutrino flavours and the geometrical factor is
 \begin{equation}\label{e:25b}
F(x)=\dfrac{f_F }{2}\left[1-g({x})\right]^2
\end{equation} 
$f_F = \sqrt{2} G_F /(2 \pi R_{\nu}^2)$ and
\begin{eqnarray}\label{e:30}
g(x)= \sqrt{1 - \left(\dfrac{R_{\nu}}{x}\right)^2 }
\end{eqnarray}
and $R_{\nu}$ the radius of the neutrinosphere. The expression for the non-linear contribution 
is explicitly
\begin{equation}\label{e:26}
G(\rho)=\sum_{\alpha=e,\mu} \int [\rho_{\nu_{\underline{\alpha}}}(q')
L_{\nu_{\underline{\alpha}}} (q')-\rho^*_{\bar{\nu}_{\underline{\alpha}}}(q')
L_{\bar{\nu}_{\underline{\alpha}}}(q')]dq'
\end{equation} 
where $L_{\nu_{\underline{\alpha}}}$  is the neutrino flux at the neutrinosphere for a neutrino "born" as an $\alpha$ flavor. 
 The second derivative of ${H}_{\nu\nu}$ includes contributions  from both the derivative of the geometrical factor and of the density matrices, i.e.
\begin{eqnarray}\label{e:27}
\ddot{H}_{\nu\nu}=\ddot{F}(x)G(\rho) +
2 \dot{F}(x)\dot{G}(\rho) + F(x)\ddot{G}(\rho)
\end{eqnarray}
To calculate $\tilde{\omega}_{\eta}$ we first need
the obvious second derivative of the geometrical term
\begin{eqnarray}\label{e:29}
\ddot{F}(x) &= &{{f_F R_{\nu}^4}\over{x^{6}}} \, 
[ g(x)^{-2} + (1-g(x))g(x)^{-3} \nonumber \\  
&&~~~~~~~~+ 3 x^2 R_{\nu}^{-2}(1-g(x))g(x)^{-1} ].
\end{eqnarray}
The second contribution to Eq.(\ref{e:27}) 
can be calculated as follows
\begin{equation}\label{e:31}
\dot{G}(\rho) = -\imath
\sum_{\alpha} \int [\left[H,\rho_{{\nu}_{\underline{\alpha}}}(q')\right] L_{{\nu}_{\underline{\alpha}}}
+  \left[\bar{H},\rho_{{\bar{\nu}}_{\underline{\alpha}}}(q')\right]^* L_{\bar{{\nu}}_{\underline{\alpha}}}]dq' 
\end{equation}
where
\begin{equation}\label{e:32}
 \dot{F}(x)= - f_{F} x^{3} R_{\nu}^{2}   \dfrac{ \ \left[1 - g(x) \right]}{g(x)}
\end{equation}
Finally the third one can be evaluated using
\begin{eqnarray}\label{e:33}
\ddot{G}(\rho)  & = & - \imath \sum_{\alpha} \int \left(\left[\dot{H},\rho_{{\nu}_{\underline{\alpha}}}(q')\right] -\imath\ \left[H,\left[H,\rho_{{\nu}_{\underline{\alpha}}}(q')\right]\right] \right)  L_{{\nu}_{\underline{\alpha}}} dq'  \\ \nonumber & & +  \left(\left[\dot{\bar{H}},\rho_{\bar{{\nu}}_{\underline{\alpha}}}(q')
\right] -\imath\ \left[\bar{H},\left[\bar{H},\rho_{\bar{{\nu}}_{\underline{\alpha}}}(q')
\right]\right] \right)^* L_{\bar{{\nu}}_{\underline{\alpha}}}dq' 
\end{eqnarray}
\section{The connection between the magnetic resonance and the spectral split phenomenon} 
\noindent
In its simplest realization, the magnetic resonance phenomenon consists in a spin-flip occurring 
in presence of a constant and a time varying magnetic field. Let us consider a spin precessing around
a constant magnetic field $\Bvec_0 = B_0 \bf{z}$, which can be taken along the z-axis, with a precession frequency given by $\omega_0$. The spin is also subject to a second magnetic field $\Bvec_1$, located in the (xOy) plane, which sinusoidally varies in time with $\omega$ frequency. Such a magnetic field
acts as a perturbation that can however significantly impact the dynamics of the system depending on
the relation between $\Delta \omega = \omega - \omega_0$ and $\omega_1$, the precession frequency around $\Bvec_1$.
If $\Delta \omega \neq 0$  the time varying $\Bvec_1$ has little effect on the spin evolution. However, if the resonance condition $\Delta \omega \sim 0 $ is met and $\Delta \omega \ll \omega_1$, the presence of $\Bvec_1$ can fully flip the spin. The amplitude for the resonant flip conversion follows a Breit-Wigner distribution $ \omega_1^2/( \omega_1^2 + \Delta \omega^2)$ which is maximal when $\Delta \omega \sim 0 $, with the resonance width being given by
$ \omega_1$. On the other hand, if $\Delta \omega$ is a multiple of $ \omega_1$ the system is off resonance while if the frequency difference is a fraction of it the system is close to resonance and the spin inversion is partial \cite{Cohen}.

\begin{figure}[h]
	\centering
		\includegraphics[width=90mm]{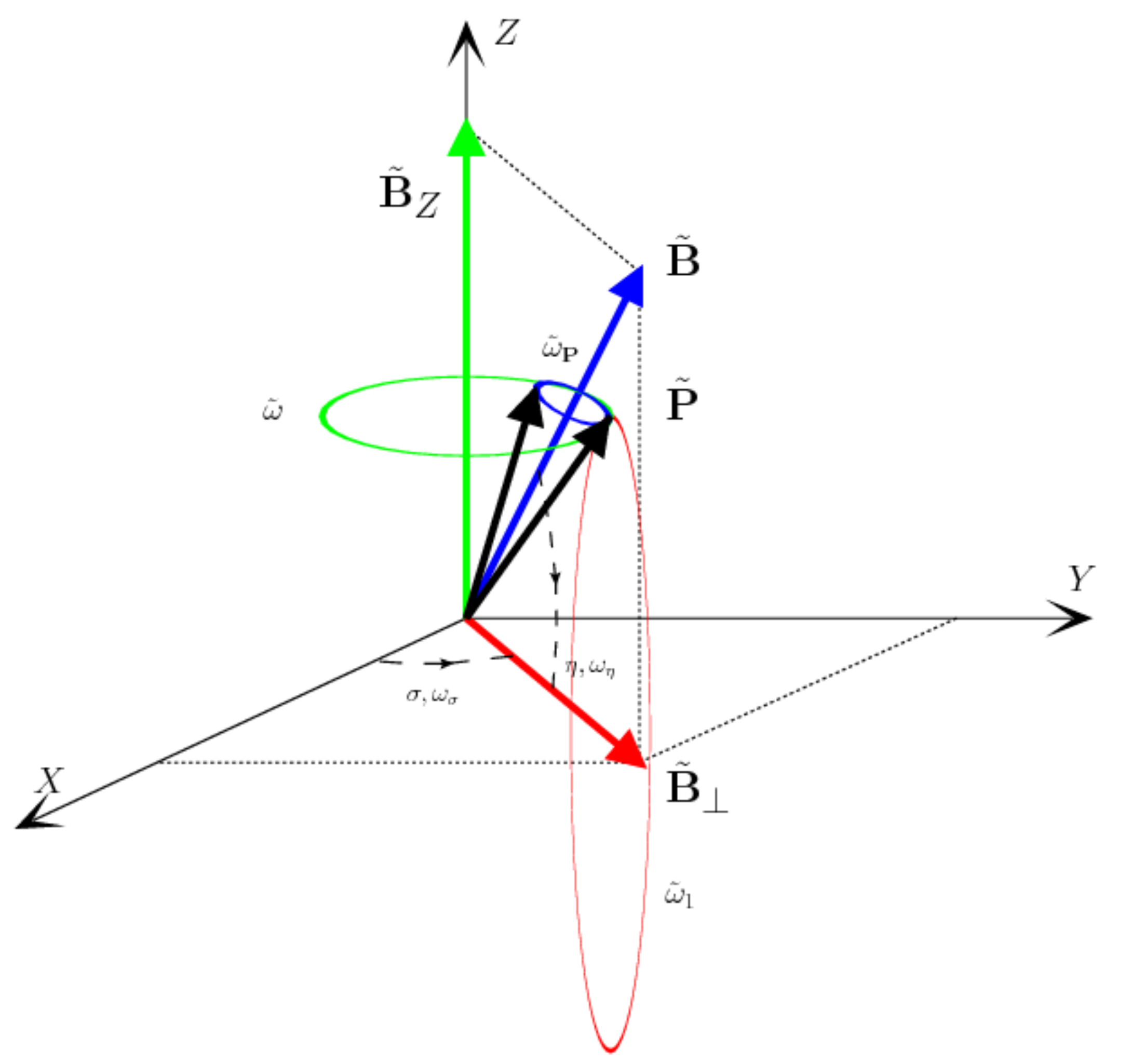}
	\caption{Analogy between the spectral split phenomenon and the magnetic resonance phenomenon, based on the polarization vector formalism. 
	The figure depicts the effective magnetic field $\tilde{\Bvec}$, its z-component $\tilde{\Bvec}_Z$ and the component lying in the (XOY) plane 
$\tilde{\Bvec}_{\perp}$	with the corresponding frequencies (see text).}
	\label{fig:polvecdiag}
\end{figure}

Let us now define three relevant frequencies that are useful to establish our connection between the neutrino spectral split phenomenon and the magnetic resonance phenomenon (Figure \ref{fig:polvecdiag}).
The effective magnetic field associated with the Hamiltonian Eq.(\ref{e:6bis}) in the flavour basis\footnote{Note that we use small $x,y,z$ (capital $X,Y,Z$) letters when in the flavour (matter) basis.} 
has components in the (xOy) plane proportional to the off-diagonal $H_{e\mu}$ matrix element which depends on the vacuum and neutrino-neutrino contributions; while the third component is generated by the matter term (that gives a large contribution) as well. 

\begin{figure}[h]
	\includegraphics[width=90mm]{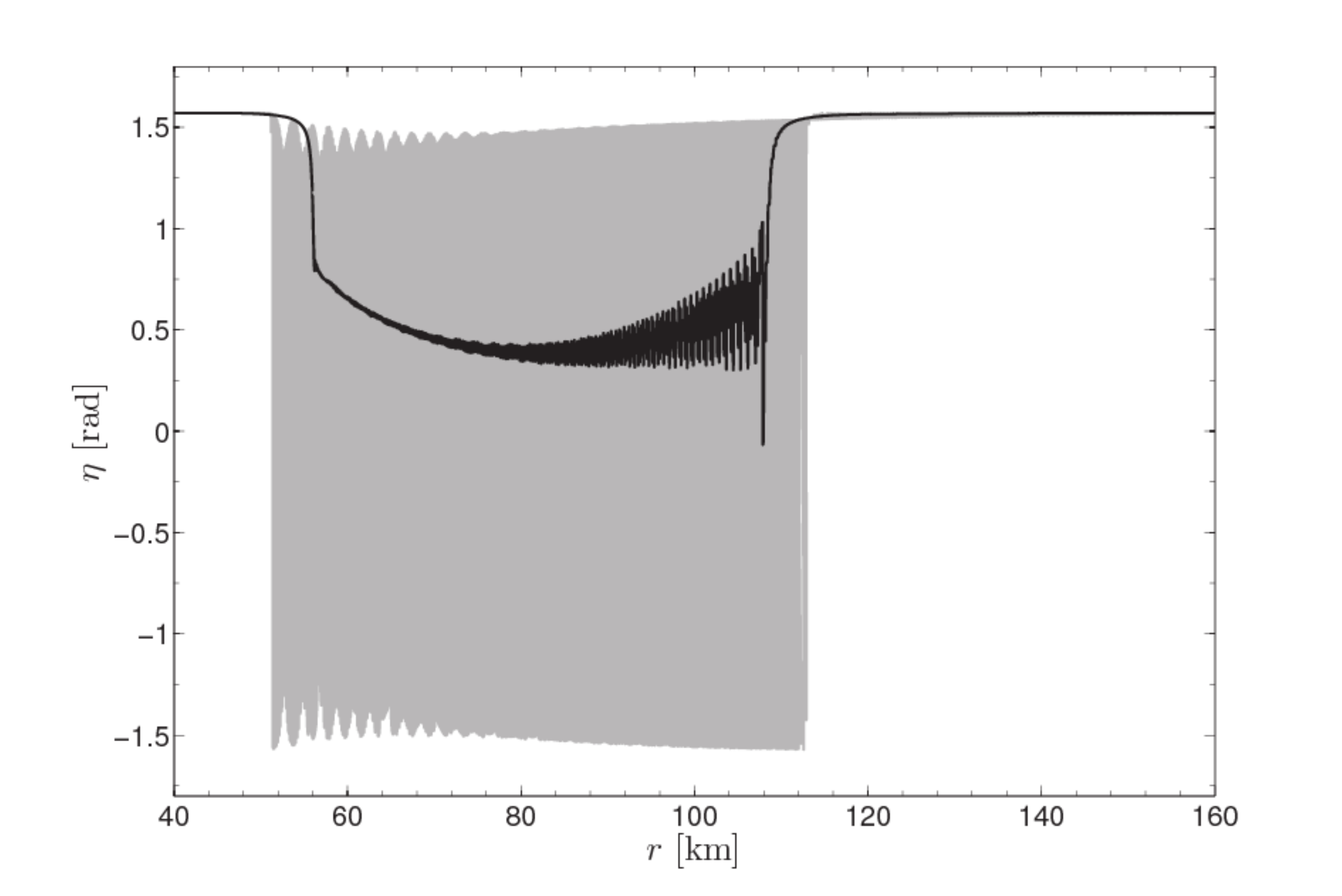}
	\caption{Time evolution of the $\eta$ angle Eq.(\ref{e:taneta}) defining the direction of the effective magnetic field $\tilde{\Bvec}$ Eq.(\ref{e:B}) with respect to the (XOY) plane (grey curve). The effective magnetic field is associated with the full matter Hamiltonian Eq.(\ref{e:Hm2}) including the vacuum contribution, the matter term and the neutrino-neutrino interaction (in the single-angle approximation).  The fast oscillations are driven by the derivative of the  $\tilde{\beta}$ phase. The black curve shows the angle with respect to the  (XOY) plane giving the evolution of  $\langle{\bf \tilde{B}} \rangle$ instead of ${\bf \tilde{B}}$ Eq. \ref{e:tanetav}. The results are obtained for a 5 MeV neutrino and inverted hierarchy.}
	\label{fig:aver1}
\end{figure}

What we would like to argue now is, that 
the inversion of the neutrino polarization vectors is due to the fact that the precession frequencies meet the magnetic resonance criteria.
To this aim, we switch to the matter basis to follow the system evolution.
Using the polarization vector formalism (Section II) we identify the $\tilde{B}_Z=B_0$ component Eq.(\ref{e:B}) for the $\Bvec_0$ magnetic field. We also identify the $\Bvec_1$ time varying magnetic field with $\tilde{\Bvec}_{\perp}$. 
We now first show that 
different flavour conversion regimes are well accounted
for, if one follows the average of the magnetic field instead of the field itself. Let us have a closer look at the time dependence of the magnetic field direction.
Figure \ref{fig:aver1} plots the fast variations of the $\eta$ angle Eq.(\ref{e:taneta}) showing that $\tilde{\Bvec}$ undergoes fast oscillations with respect to the (XOY) plane.
The numerical results presented in this work have been obtained by using the vacuum oscillation parameters $|\delta m^2|= 2.4 \times 10^{-3}$eV$^2$, $\theta=9^\circ$ and 
the matter density profile 
$\rho_B=1.5 \times 10^8\,(10/x)$, with $x$ being here the distance in the supernova, in units of g.cm$^{-3}$ and of km (for $x$).
The neutrinosphere $R_{\nu}$ is taken at 10 km and the corresponding neutrino fluxes are assumed to be of Fermi-Dirac type with average energies  $ \langle E_{\nu_e} \rangle =12$ MeV, $ \langle E_{\bar{\nu}_e}\rangle=15$ MeV and $ \langle E_{\nu_{x}} \rangle=18$ MeV. Equipartition of energy is assumed, with a total luminosity of $4\times10^{51}$ erg.s$^{-1}$,
While we present one set of our numerical results, we stress that we have performed calculations for an ensemble of initial conditions, varying the neutrino average energies and luminosities at the neutrinosphere. The quality of the results we obtain is the same for all initial conditions considered, so that we show here one case only. 

Our calculations of the ratio of the $ \tilde{\omega}_\eta$ Eq.(\ref{e:omegaeta}) and $ \tilde{\omega}_{\bf P}$ Eq.(\ref{e:omegaP})
frequencies, i.e. $ \tilde{\omega}_\eta/\tilde{\omega}_{\bf P}$ shows
that such ratio is always much larger than 1 : the 
magnetic field direction changes rapidly compared to the precession of the neutrino polarization vector around it. 
Therefore let us see how well the system is described
if we replace $\tilde{\Bvec}$ by $\langle \tilde{\Bvec} \rangle$. To do such a comparison, in order to obtain our  $\langle \tilde{\Bvec} \rangle$, we perform a numerical 
average of the magnetic field $\tilde{\Bvec}$ over 10 km.
If we define the $\sigma $ angle describing the $\tilde{\Bvec}_{\perp}$ evolution in the (XOY) plane,
\begin{equation}\label{e:tans}
 \tan \sigma = \dfrac{ \langle \tilde{\Bvec}_Y  \rangle}{ \langle \tilde{\Bvec}_X  \rangle} = \dfrac{\mathcal{R}_{av}({\Gamma})}{\mathcal{I}_{av}({\Gamma})}
\end{equation}
where the subscript ${av}$ means that we are taking the average of the corresponding quantity 
$\mathcal{R}_{av}({\Gamma}) = \langle \mathcal{R}({\Gamma}) \rangle$ and $\mathcal{I}_{av}({\Gamma}) = \langle \mathcal{I}({\Gamma}) \rangle$
and the associated angular velocity,
\begin{equation}\label{e:omegas}
 \tilde{\omega}_\sigma = \dot{\sigma} = \dfrac{1}{|\Gamma_{av}|^2} \left(\mathcal{I}_{av}({\Gamma}) \mathcal{R}_{av}({\dot{\Gamma}})- \mathcal{R}_{av}({\Gamma}) \mathcal{I}_{av}(\dot{\Gamma})\right) 
\end{equation}
The $\tilde{\omega}_\sigma$ frequency involves the derivative of the $\Gamma$ factor Eq.(\ref{e:derGamma}). In our numerical calculations we obtain that the $\mathcal{R}_{av}({\dot{\Gamma}})$, $\mathcal{R}_{av}({\Gamma})$ and therefore the $\tilde{\omega}_\sigma$ frequency are very close to zero.
The second quantity of interest is the precession frequency around 
$\tilde{B}_Z$, namely $\tilde{\omega} = |\langle \tilde{B}_Z \rangle|= |\delta \tilde{k}_{12} - \dot{\tilde{\beta}}\rm{cos}2 \tilde{\theta} |$. 
Another relevant precession frequency is the one around the $\tilde{\Bvec}_1$,
namely 
\begin{equation}\label{e:om1}
\tilde{\omega}_1 = |\tilde{\Bvec}^{av}_{\perp}| = \sqrt{\langle \tilde{B}_X \rangle^2 + \langle \tilde{B}_Y \rangle^2} =
\sqrt{ \langle 
(\dot{\tilde{\beta}}\rm{sin}2 \tilde{\theta})^2 \rangle  + \langle 4 \dot{\tilde{\theta}}^2 \rangle }
\end{equation} 
Since the contribution coming from $\langle 4 \dot{\tilde{\theta}}^2 \rangle$ comes out to be small, the quantity $\tilde{\omega}_1$ is essentially determined by the phase derivative.
Finally one can also define the same quantities as in Eqs.(\ref{e:omegaP}) and (\ref{e:omegaeta}) but this time for the average magnetic field, namely
\begin{equation}\label{e:omPav}
\tilde{\omega}_{\bf P} = {\bf | \langle \tilde{B} \rangle  |}.
\end{equation} 
The rate of change of the angle defining the direction of $\langle{\bf \tilde{B}} \rangle$ with 
respect to the (XOY) plane is now given by
\begin{equation}\label{e:tanetav}
\tan\eta = \dfrac{\langle \tilde{\Bvec}_Z \rangle}{ \left|  \tilde{\Bvec}_\perp^{av}  \right|} = \dfrac{ \langle \dBk{12} \rangle}{\left| \langle 2 \tilde{H}_{12}   \rangle \right|}
\end{equation}
with $ \tilde{\Bvec}_\perp^{av} = (\langle \tilde{B}_X \rangle, \langle \tilde{B}_Y \rangle)$.
We show in Figure  \ref{fig:aver1} an example of the evolution
of the $\eta$ angle defining the evolution of $\langle{\bf \tilde{B}} \rangle$ given by Eq.(\ref{e:tanetav}) for a 5 MeV neutrino (black curve),
instead of ${\bf \tilde{B}}$ (grey curve).

\begin{figure}[h]
		\includegraphics[width=90mm]{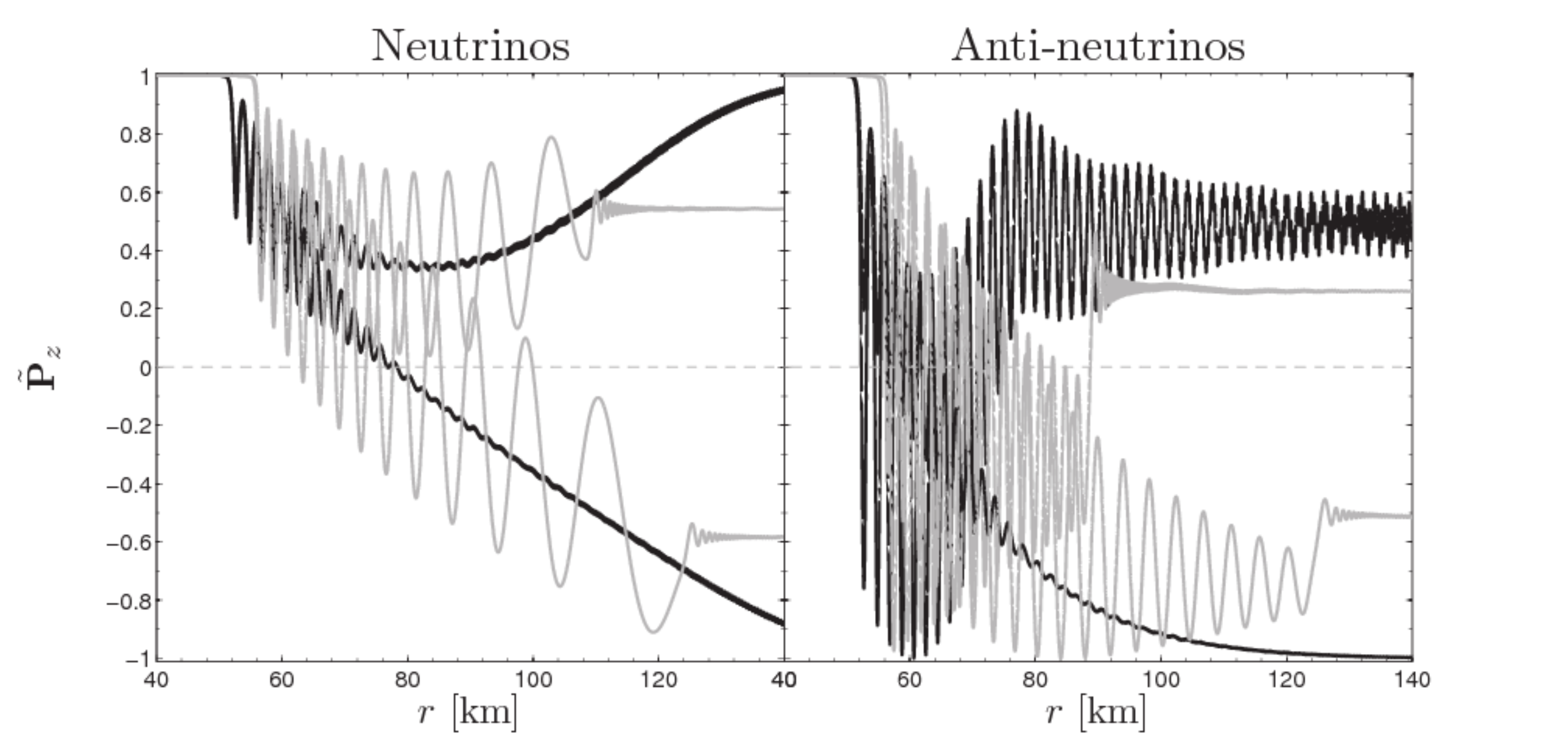}
	\caption{Evolution of the third component of the polarization
	vector Eq.(\ref{e:P}) as a function of distance within a core-collapse supernova, determined by using the magnetic field $\tilde{\Bvec}$ Eq.(\ref{e:B}) (black curves) or its average $\langle \tilde{\Bvec} \rangle$ (gray curves). Results are shown for a 5 and a 10 MeV neutrino (left-figure) and for a 5 and a 2 MeV antineutrino (right figure). The numerical calculations include the full matter Hamiltonian Eq.(\ref{e:Hm2}) with vacuum mixings, neutrino coupling to matter and to neutrinos (in the single-angle approximation). The neutrino energy dependent behaviour characteristic of the spectral split phenomenon (giving rise to a complete flavour conversion and a spectral swap) can be seen. }
	\label{fig:polav}
\end{figure}

We have therefore numerically verified, that the evolution of our system is well described if one uses the $\langle
\tilde{\Bvec}  \rangle = (\langle \tilde{B}_X \rangle, \langle \tilde{B}_Y \rangle, \langle \tilde{B}_Z \rangle)$ 
instead of the magnetic field itself. To this aim we have performed two calculations for the neutrino evolution 
and determined the corresponding 
oscillation probabilities. In the first calculation we have solved the precession equation Eq.(\ref{e:precession})  with $\tilde{\Bvec}$. In the second we have 
determined $\langle \tilde{\Bvec} \rangle$ making a numerical average over 10 km (as mentioned above) and then solved Eq.(\ref{e:precession}) using  the average of the effective magnetic field instead of the field itself. Figure \ref{fig:polav} presents the comparison of the survival 
$P(\nu_i \rightarrow \nu_i)$ probabilities with $i=1,2$
for the matter eigenstates, obtained when the neutrino evolution is determined by either
$\tilde{\Bvec}$ or $\langle \tilde{\Bvec} \rangle$. The system evolves
 near the neutrinosphere where the neutrino-neutrino interaction is important, with the evolution determined by
the full matter Hamiltonian  Eq.(\ref{e:Hm2}).
Results are shown both for neutrinos and for anti-neutrinos of two different energies. One can see that the synchronization, bipolar and spectral split
regimes can be easily identified in both cases. Note that the different behaviours of the probabilities shown in Figure \ref{fig:polav} engender the characteristic spectral swap of the electron with the other $\nu_x$ flavour fluxes (Figure \ref{fig:flux}).
One can see that, when averaging the fast variations due to 
$\dot{\tilde{\beta}}$, the main features of the synchronized, the bipolar and the spectral split behaviours remain unchanged.
Note that the two descriptions agree qualitatively and quantitatively very well until 120 km for the neutrino, and 80 km for the anti-neutrino cases. Afterwards the evolution of the systems following the average of $\tilde{\Bvec}$ and $\tilde{\Bvec}$ itself start presenting some differences.
However, looking carefully at the survival probabilities both in the matter and in the flavour basis (Figure \ref{fig:Gamma}), one can see
that
the region where the two descriptions compare well is sufficient for our purposes since discrepancies appear at the very end of the spectral split behaviour in all cases. This convince us that the results we will present in the following are not an artefact of following the average of the effective magnetic field. Therefore, from now on,
we will replace $\tilde{\Bvec}$ by $\langle \tilde{\Bvec} \rangle$. All the results of the next section are obtained following such a prescription. 

\section{The occurrence of the spectral split and the magnetic resonance condition}
The main goal here is to show that the magnetic resonance conditions :
\begin{equation}\label{e:res}
\begin{array}{c}
\Delta\omega = \omega-\omega_0 = 0 \\
\dfrac{\Delta\omega}{\omega_1}\ll 1
\end{array}
\quad\Leftrightarrow\quad
\begin{array}{c}
\tilde{\omega}=0 \\
\dfrac{\tilde{\omega}}{\tilde{\omega}_1}\ll 1
\end{array}
\end{equation}
are verified for the neutrino or anti-neutrino energies and at the same location in the supernova 
for which the spectral split occurs. We emphasize that verifying that the spectral split is a magnetic 
resonance phenomenon can be done both in the frame identified by the flavour basis and in the one given 
by the comoving frame. 
In the flavor basis  the effective magnetic field has indeed a small time varying component $\Bvec_{\perp}$ in the (xOy) plane; the magnetic resonance conditions are those given on the left of (\ref{e:res}).
On the other hand, in the frame identified by the matter basis and the average of $\tilde{\Bvec}$,
$\langle \tilde{\Bvec}_{\perp} \rangle$ appears as static and the magnetic resonance conditions are those on the right side of 
  (\ref{e:res}).
For convenience we show the fulfillement of the magnetic resonance conditions in the comoving frame identified by the average of the magnetic field in the matter basis, i.e.  $\langle \tilde{\Bvec} \rangle$. 

Let us follow the different phases of the neutrino evolution with
the polarization vector formalism in the matter basis. At early times, near the neutrinosphere, the matter term is so large that the polarization vector evolution is essentially determined $\langle \tilde{\Bvec}_Z \rangle \approx \langle \delta \tilde{k}_{12} \rangle$. 
At this stage it is essentially precessing around this component and nothing happens from the point of view of flavour conversion.
As shown in \cite{Galais:2011jh}, at the end of the synchronization regime,
the onset of bipolar oscillations is triggered by a rapid growth of
the $\dot{\tilde{\beta}}$ phase. 
This implies that the $\tilde{\Bvec}^{av}_{^\perp}$ components -- related to the off-diagonal elements of the matter Hamiltonian -- start being important, as does the phase contribution to  $\langle \tilde{\Bvec}_Z \rangle$. 
In fact the phase growth introduces an oscillating degeneracy between the off-diagonal Hamiltonian matrix element and the difference of the diagonal matrix elements.
As a consequence the ratio
of such elements determining the generalized adiabaticity parameter, become comparable, 
as discussed in \cite{Galais:2011jh}. In the present formalism such parameters are given by cot$\eta$ (see Eq.(\ref{e:tanetav})) :
\be\label{e:Gamma} 
\Gamma = {{ |\tilde{\Bvec}^{av}_{^\perp} | }\over{ \left| \langle  \tilde{\Bvec}_Z \rangle  \right|}} = {{|\langle 2 \tilde{H}_{12} \rangle |}\over {|\langle \delta \tilde{k}_{21} + \delta Q_{21}  \rangle |}} .
\ee

\begin{figure}[h]
	\centering
		\includegraphics[width=90mm]{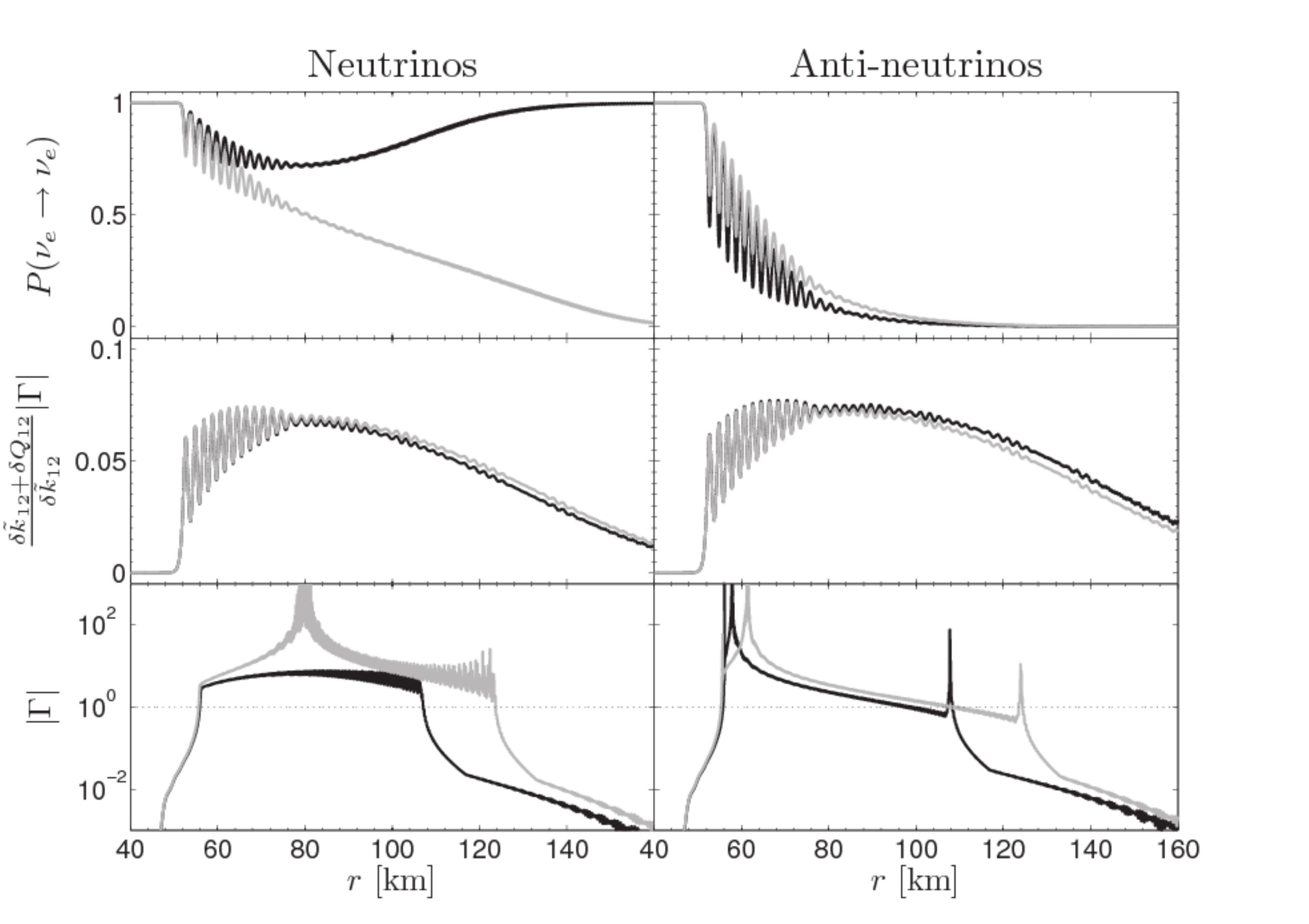}
	\caption{Evolution of survival probabilities of the flavour eigenstates (upper), of the corresponding generalized adiabaticity parameters $\Gamma$ Eq.(\ref{e:Gamma}) (lower) and of the same quantity multiplied by the factor $(\delta \tilde{k}_{12} + \delta Q_{12})/\delta \tilde{k}_{12}$ (middle figures). The results are obtained by solving Eq.(\ref{e:Hm2f}) for the full Hamiltonian comprising the vacuum contribution, the matter term and the neutrino-neutrino interaction (in the single-angle approximation). The case is inverted hierarchy. In each figure the two curves correspond to different neutrino energies, namely a 5 MeV (black) and a 10 MeV (gray) neutrino or anti-neutrino. }
	\label{fig:Gamma}
\end{figure}

\begin{figure}[h]
	\centering 
		\includegraphics[width=90mm]{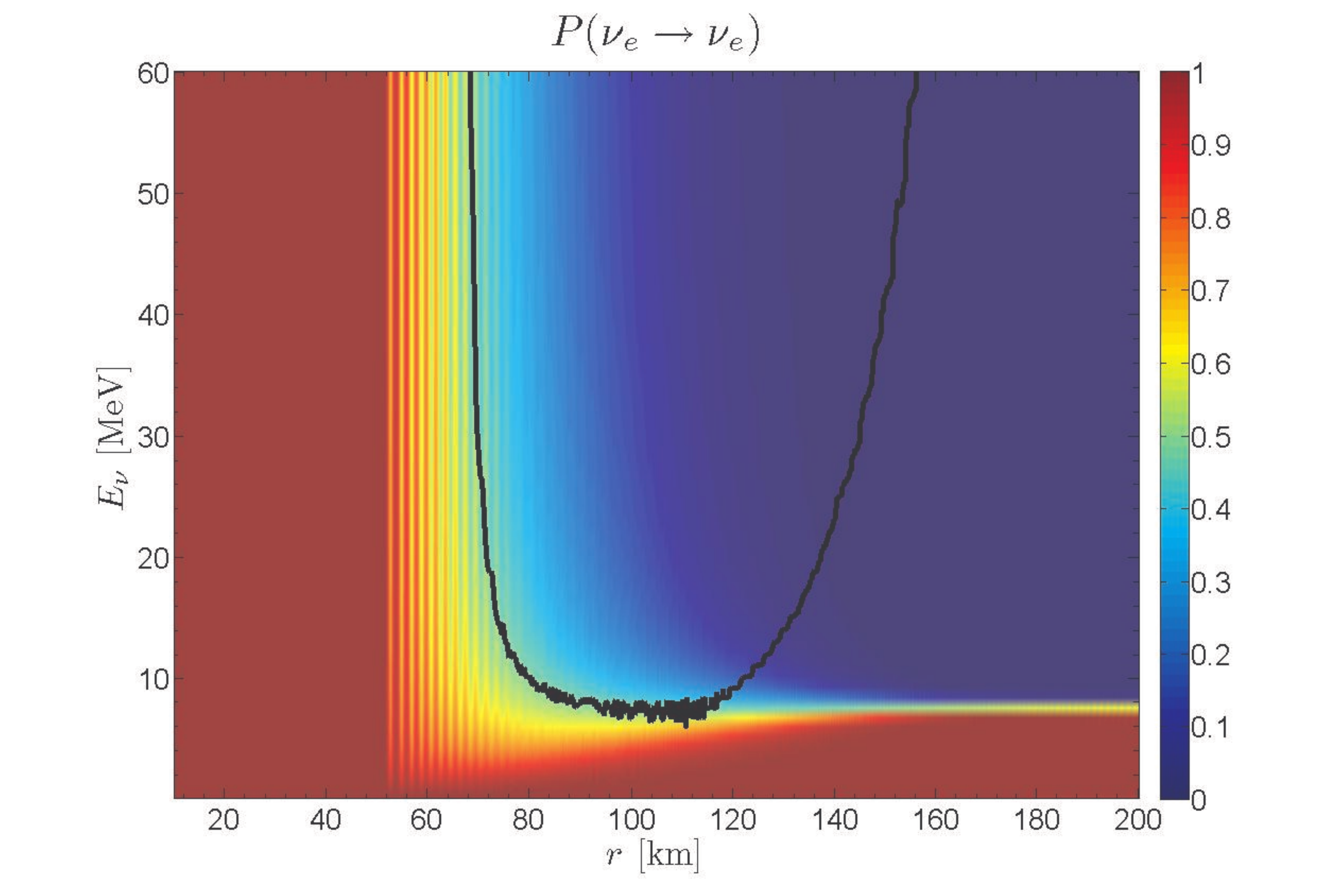}
	\caption{Three dimensional contour 
	plot of the electron neutrino survival probability (in the flavour basis) as a function the neutrino energy and distance within the
	supernova. The colors encode the electron neutrino survival probability. The yellow-green region corresponds to a survival probability of 0.5 so that all neutrino energies below the split energy of 7.6 MeV do not undergo any spectral swap, while all energies above the split energy undergo a spectral swap. The black curve superposed shows when the magnetic
 resonance condition in the matter basis is met : $\tilde{\omega}=0$ Eq.(\ref{e:res}) (see Figure \ref{fig:nmrnu}).}
	\label{fig:Pee}
\end{figure}

\begin{figure}[h]
	\centering
		\includegraphics[width=90mm]{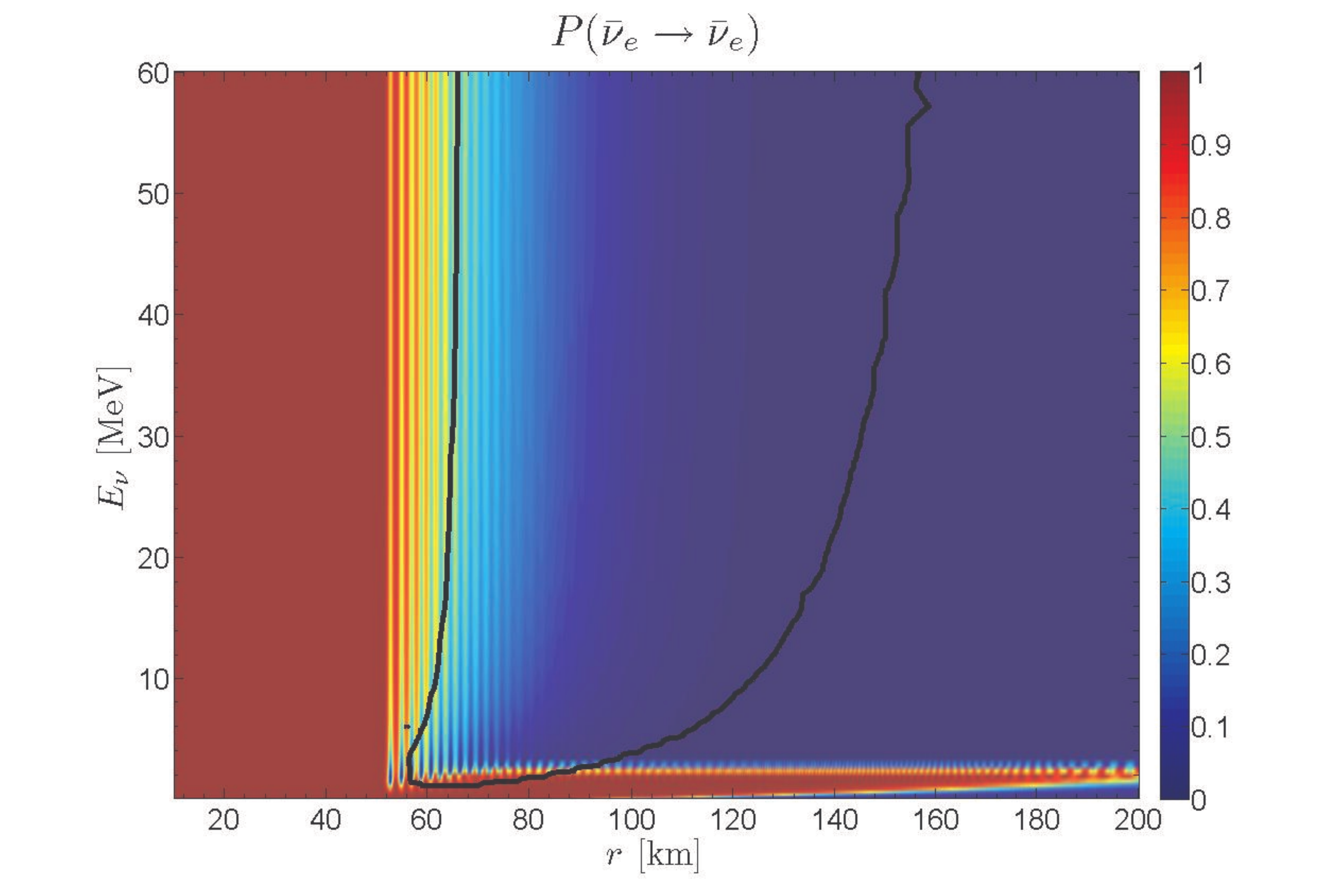}
	\caption{Three dimensional contour 
	plot of the electron anti-neutrino survival probability (in the flavour basis) as a function the anti-neutrino energy and distance within the
	supernova. The colors encode the electron anti-neutrino survival probability. The yellow-green region corresponds to a survival probability of 0.5 so that all neutrino energies below the split energy of 2.3 MeV do not undergo any spectral swap, while all energies above the split energy undergo a spectral swap. The black curve superposed shows when the 
magnetic resonance condition in the matter basis is met : $\tilde{\omega}=0$ Eq.(\ref{e:res}) (see Figure \ref{fig:nmranu}).}
	\label{fig:Peea}
\end{figure}

\begin{figure}[h]
	\centering
	\includegraphics[width=90mm]{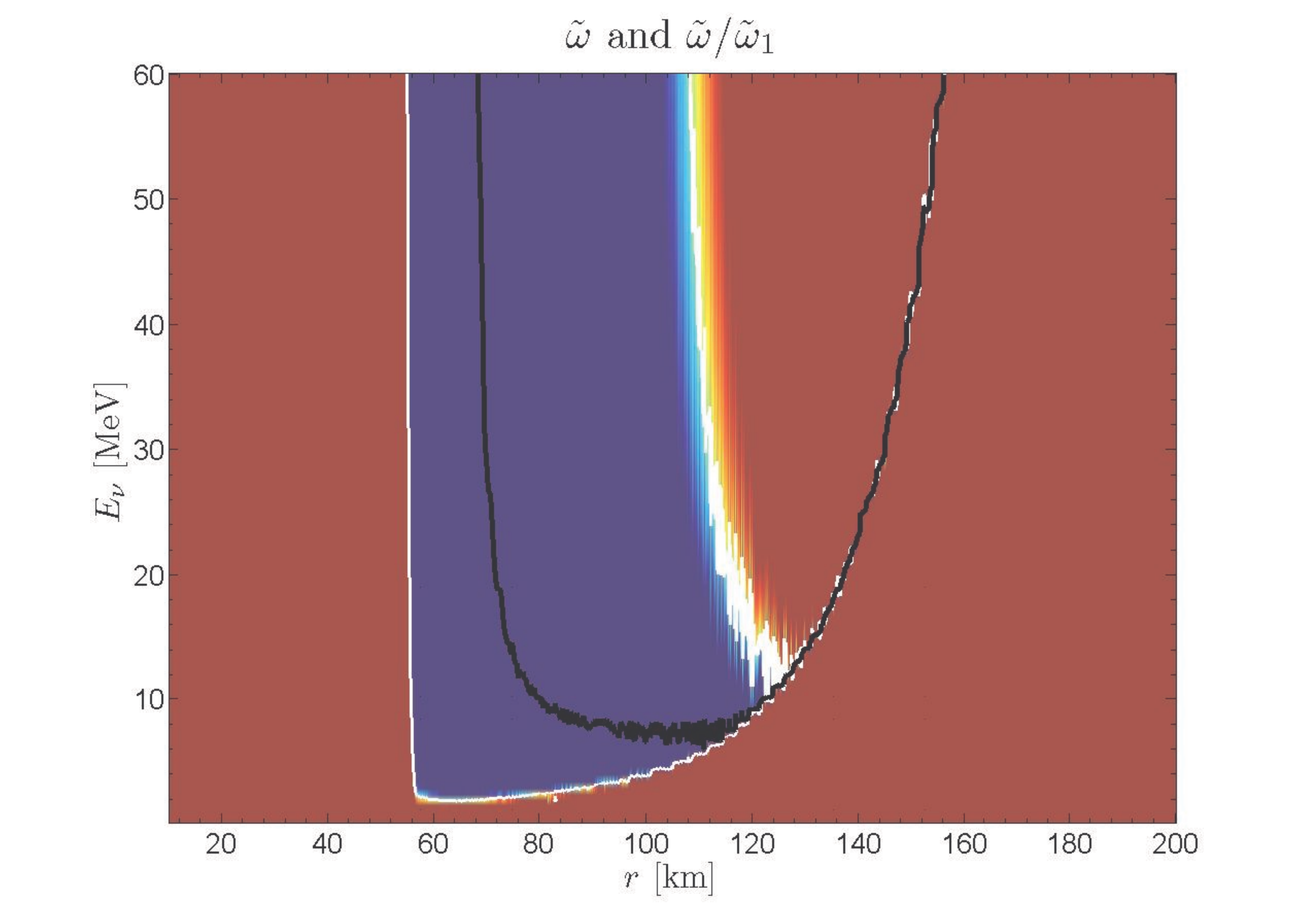}
	\caption{Connection of the spectral split phenomenon with the magnetic resonance phenomenon for neutrinos: 
The colored blue region shows the neutrino energies and the supernova location for which the resonance conditions
$\tilde{\omega} \approx 0$ and $\tilde{\omega}/\tilde{\omega}_1 \ll 1$ (on resonance) Eq.(\ref{e:res}) 
are both fulfilled. 
The black curve presents the location of the magnetic resonance corresponding to the fulfillment of the 
$\tilde{\omega} \approx 0$ condition. The white line separates the area where we are on resonance from
the one off-resonance given by $\tilde{\omega}/\tilde{\omega}_1 \gg 1$.
Note that the quantity $\omega_1$ gives the magnetic resonance width. 
The figure shows that the resonance criteria follow a line, one around 75 km where the flipping of 
the neutrino flavour polarization vector occurs, that then stays flat and then turns upward at around 
120 km. Note that in this second region the polarization vectors are not flipped since the criteria 
$\tilde{\omega}/\tilde{\omega}_1 \ll 1$ is not met anymore (the neutrino-neutrino 
interaction has become negligible).}
	\label{fig:nmrnu}
\end{figure}

\begin{figure}[h]
	\centering
		\includegraphics[width=90mm]{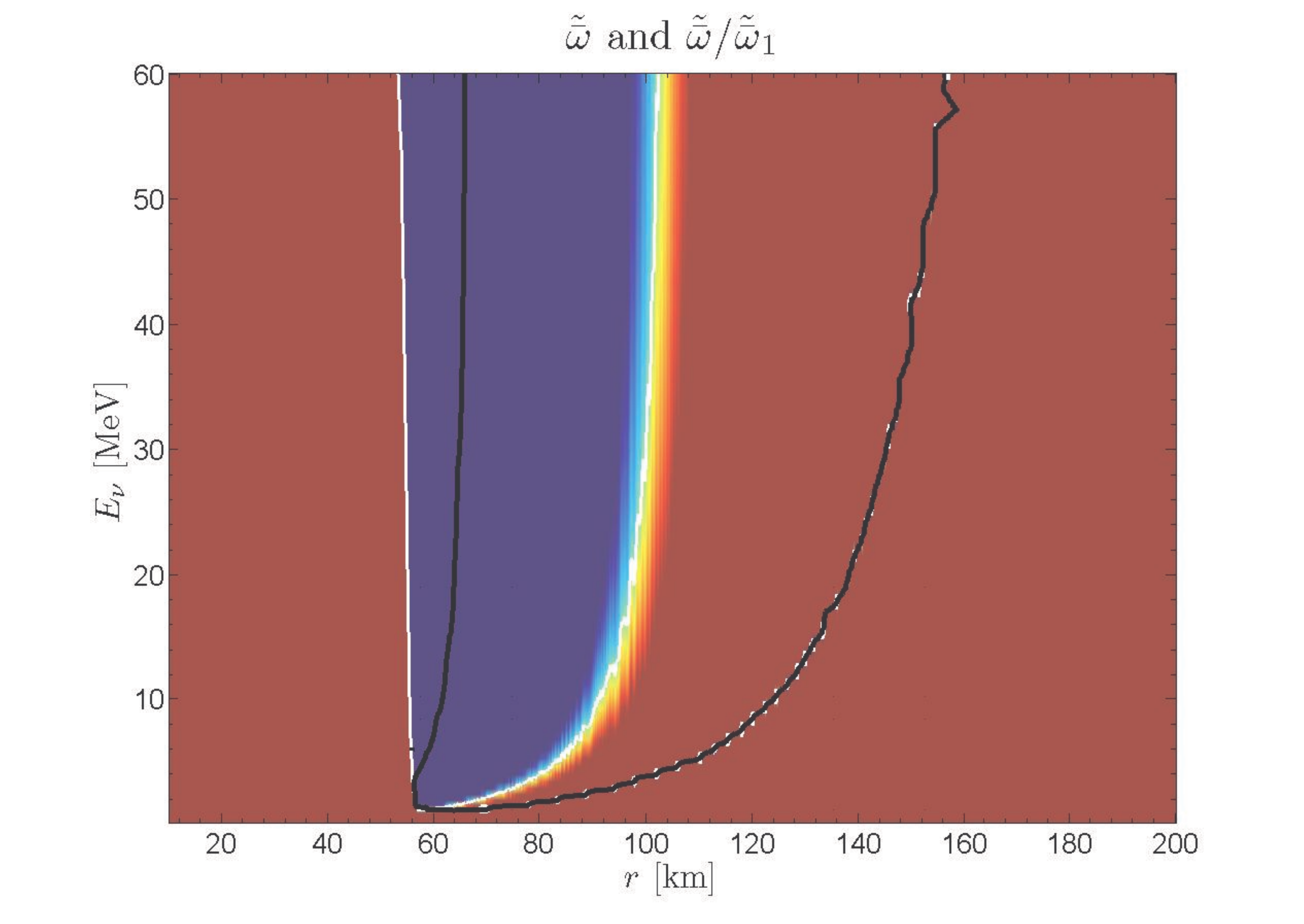}
	\caption{Connection of the spectral split phenomenon with the magnetic resonance phenomenon for neutrinos: 
The colored blue region shows the neutrino energies and the supernova location for which the resonance conditions
$\tilde{\omega} \approx 0$ and $\tilde{\omega}/\tilde{\omega}_1 \ll 1$ (on resonance) Eq.(\ref{e:res}) 
are both fulfilled. 
The black curve presents the location of the magnetic resonance corresponding to the fulfillment of the 
$\tilde{\omega} \approx 0$ condition. The white line separates the area where we are on resonance from
the one off-resonance given by $\tilde{\omega}/\tilde{\omega}_1 \gg 1$.
Note that the quantity $\omega_1$ gives the magnetic resonance width. 
The figure shows that the resonance criteria follow a line, one around 65 km where the flipping of the neutrino flavour polarization vector occurs, that then stays flat and then turns upward at around 100 km. Note that in this second region the polarization vectors are not flipped since the neutrino-neutrino interaction has become negligible.} 
	\label{fig:nmranu}
\end{figure}

During the bipolar oscillations the polarization vector evolution is determined essentially by the two components $(\tilde{\Bvec}_X, \tilde{\Bvec}_Y)$. However when taking the average, since
the $\dot{\tilde{\theta}}$ average is around zero (note that the derivative
of angle turns out to be much smaller than the phase derivative), the effective magnetic field associated with the matter Hamiltonian
stays in a XZ plane. 
This is the co-rotating frame already discussed in the literature \cite{Duan:2005cp,Duan:2007fw,Raffelt:2007cb}. After some distance the polarization vector enters in a new phase of its evolution : the magnetic resonance regime. 
In this phase, depending on the neutrino energy, if the resonance condition $\Delta \omega = \omega - \omega_0 \approx 0$ 
is satisfied then a full inversion of the neutrino polarization vector occurs. In the comoving frame in the matter basis such a condition
reads\footnote{Note that in the corotating frame $\tilde{\omega}=\omega - \omega_0$.} $\tilde{\omega} \approx 0$ and near resonance $\tilde{\omega}/\tilde{\omega}_1 \ll 1$. 

Figure \ref{fig:Gamma} presents the evolution, in the flavour basis, of survival probabilities for neutrinos and anti-neutrinos in inverted hierarchy, in correspondence with the $|\Gamma|$ evolution. 
The latter quantity, multiplied by the factor 
$(\delta \tilde{k}_{12} + \delta Q_{12})/\delta \tilde{k}_{12} $, 
is also shown to emphasize the role of the $\delta Q_{12}$, coming from the $\TU^{\dagger}{d\TU}/{dx}$ contribution to the diagonal elements of the matter Hamiltonian. Results are presented for two different neutrino 
and anti-neutrino energies chosen to show the different behaviour of the generalized adiabaticity parameters for neutrinos that do not (do) undergo the spectral split phenomenon and a swapping of the fluxes depending on their energy. One can see from the figure that the corresponding
 $|\Gamma|$ has a very different behaviour in presence of the swaps (i.e. the anti-neutrino and the 10 MeV neutrino cases) with respect to the absence (the 5 MeV neutrino case). In particular, for the former cases at the location of the resonance spikes appear in the adiabaticity parameter evolution, at 80 km for the 10 MeV neutrino, and around 60 km for the anti-neutrinos. Note that the second spike at 110 (130) km for a 5 (10) MeV anti-neutrino is not associated with any significant change because the neutrino-neutrino interaction term has become negligible and the second condition
$\tilde{\omega}/\tilde{\omega}_1 \ll 1$ is not satisfied (see Figs. \ref{fig:nmrnu} and \ref{fig:nmranu}) .

Figures \ref{fig:Pee} (neutrino case) and \ref{fig:Peea} (anti-neutrino case) present three dimensional contour plots of the neutrino survival probabilities in the flavour basis for different neutrino energies, as a function of distance within the supernova. 
Figures \ref{fig:nmrnu} and \ref{fig:nmranu} show the regions where the resonance $\tilde{\omega} \approx 0$ (black lines) and near resonance $\tilde{\omega}/\tilde{\omega}_1 \ll 1$ (band within white curves) conditions are met  in the matter basis. The coloured area outside the white shaded region
is a region off-resonance $\tilde{\omega}/\tilde{\omega}_1 \gg 1$.
By comparing Figure \ref{fig:Pee} with \ref{fig:nmrnu} and Figure \ref{fig:Peea} with \ref{fig:nmranu} one sees that : i) the same neutrino and anti-neutrino energies for which the magnetic resonance condition is fulfilled also undergo a spectral swap; ii) the inversion of the neutrino flavour polarization vector occurs at the same distance in the supernova where the resonance occurs. One can also see that the resonance region in the anti-neutrino case is much smaller than in the neutrino case in agreement with the faster decrease of the anti-neutrino survival (or appearance) probabilities in the flavour and matter basis, compared to the neutrino case (see Figure \ref{fig:Gamma}). 
Finally the upturn of the black and white curves in Figures \ref{fig:nmrnu} and \ref{fig:nmranu} shows a region where the condition $\tilde{\omega} \approx 0$ is fulfilled while $\tilde{\omega}/\tilde{\omega}_1 \approx 1$. No inversion of the neutrino flavour polarization vectors is observed in such a case, in correspondence with the second spike observable in the adiabaticity parameters $\Gamma$ (Figure \ref{fig:Gamma}). 
When the precession frequencies are way out of resonance and the role of the neutrino-neutrino interaction clearly negligible, the system  evolution is determined again by the $\tilde{\Bvec}_Z$ component, given by $\delta \tilde k_{12} $. 

\section{Conclusions}
We have investigated collective flavour conversion effects, within two-flavours, induced by the neutrino-neutrino interaction in a core-collapse supernova. Using the neutrino polarization vector formalism associated with the matter Hamiltonian and the neutrino matter basis, we have focussed upon the spectral split phenomenon and established a connection with a magnetic resonance phenomenon. In such a formalism, the third component of the effective magnetic field is given by the difference of the diagonal matrix elements of the matter Hamiltonian comprising the matter eigenvalues and the phase derivative induced by the presence of complex off-diagonal terms in the neutrino flavour Hamiltonian. The X- and Y-components of the effective magnetic field depend upon the derivative of the matter phase and angle respectively. By defining precession frequencies around the average of the Z-component and of these perpendicular components, our numerical calculations of the polarization vector evolution have allowed the identification of a new phase in the evolution : the magnetic resonant regime. This is encountered when the magnetic resonant criteria concerning the precession frequencies is met. Our results show this very interesting feature : the neutrino energies for which the resonant criteria are fulfilled are the same energies undergoing the spectral split phenomenon, occurring at the same location
in the supernova. When the system is in such a regime the corresponding generalised adiabaticity parameters present spikes, while they present a smooth behaviour for the neutrino energies that do not meet the resonant  and do not undergo any spectral swap. Once the connection between the spectral split and a magnetic resonance phenomenon established, it is clear that the fulfillment of the magnetic resonance conditions can be done either in the flavour basis, or in the comoving frame (identified here by the average of the magnetic field in the matter basis).  While we have used the latter for convenience, picking one or the other is, obviously, just a question of choice.    

In this work the numerical results have been obtained assuming
equipartition of neutrino energies at the neutrinosphere and inverted 
hierarchy since for this case the neutrino-neutrino interaction gives rise to
single splits. 
While the results presented in this work correspond to a set of initial conditions in terms of neutrino average energies and luminosities at the neutrinosphere, results of the same quality are obtained for a large set of initial conditions.
We think the mechanism we have been identifying to be general and valid also in the case of multiple splits and of three neutrino flavours. Such investigations will be the object of future work.

\section{Appendix}
Here we give expressions for
the second derivative of the neutrino-neutrino interaction Hamiltonian valid in the case of a multi-angle treatment. In this case the non-linear Hamiltonian reads
\begin{eqnarray}\label{e:29b}
H_{\nu\nu}^{m.a.}&=&\dfrac{\sqrt{2} G_F}{2 \pi R_{\nu}^2}
 \sum_{\alpha} \int^{\infty}_{0} \int^{1}_{\cos\theta_{max}} dq'\,d\cos\theta'\left(1-\cos\theta \cos\theta'\right) \nonumber\\
&&[\rho_{{\nu}_{\underline{\alpha}}} (q',\theta') L_{{\nu}_{\underline{\alpha}}}
(q') - \rho_{\bar{{\nu}}_{\underline{\alpha}}}^*(q',\theta') L_{\bar{{\nu}}_{\underline{\alpha}}}(q')]
\end{eqnarray}
with
\begin{equation}
 \cos\theta_{max}=\sqrt{1-\left(R_\nu/x\right)^2}
\end{equation}
Using the Leibniz Integral Rule :
\begin{eqnarray}\label{e:30b}
\dfrac{\partial}{\partial z} \int^{b(z)}_{a(z)} f(y,z) dx &=& \int^{b(z)}_{a(z)} \dfrac{\partial f(y,z)}{\partial z} dx \nonumber\\
&&+ f(b(z),z) \dfrac{\partial b(z)}{\partial z} \nonumber\\
&&- f(a(z),z) \dfrac{\partial a(z)}{\partial z}
\end{eqnarray}
one can show that the derivative for Eq.(\ref{e:29b}) is \cite{Galais:2011jh} 
\begin{equation}
 \dot{H}_{\nu\nu}^{m.a.} = \dfrac{\sqrt{2} G_F}{2 \pi R_{\nu}^2} \sum_{\alpha} \int^{\infty}_{0}dq'\,
 \left( R(\theta',q')- S(x)\,T(q') \right)
\end{equation}
with
\begin{subequations}
 \begin{eqnarray}\label{e:31b}
R(\theta',q')&=& \int^{1}_{\cos\theta_{max}}d\cos\theta'\,\left(- \imath \left(1-\cos\theta \cos\theta'\right)\right. \nonumber\\
&&\left[\left[ H, \rho_{{\nu}_{\underline{\alpha}}}(q',\theta')\right] L_{{\nu}_{\underline{\alpha}}}(q') \right. \nonumber\\
&&\left.\left. + \left[ \bar{H}, \rho_{{\bar{\nu}}_{\underline{\alpha}}}(q',\theta')\right]^* L_{\bar{{\nu}}_{\underline{\alpha}}}(q')\right]\right) \\
S(x)&=& \dfrac{\left(R_\nu /x\right)^2}{x\sqrt{1-\left(R_\nu/x\right)^2}}\\
T(q')&=& \left(1-\cos\theta \cos\theta_{max}\right)[\rho_{\nu_{\underline{\alpha}}} (q',\theta_{max}) L_{\nu_{\underline{\alpha}}}(q') \nonumber\\
&& - \rho_{\bar{{\nu}}_{\underline{\alpha}}}^*(q',\theta_{max}) L_{\bar{{\nu}}_{\underline{\alpha}}}(q')]
\end{eqnarray} 
\end{subequations}

The second derivative of ${H}_{\nu\nu}^{(f)}$ is given by:
\begin{eqnarray}
 \ddot{H}_{\nu\nu}^{(f)} & = & \dfrac{\sqrt{2} G_F}{2 \pi R_{\nu}^2} \sum_{\alpha} \int^{\infty}_{0}dq'\,
 \left( \dot{R}(\theta',q') \right. \nonumber\\
 & & \left.\qquad- \dot{S}(x)\,T(q') - S(x)\,\dot{T}(q') \right)
\end{eqnarray}
Its evaluation requires the first derivatives of the 
$R$, $T$ and $S$ functions.
For $\dot{R}$ we get :
\begin{equation}
 \dot{R}(\theta',q')= \int^{1}_{\cos\theta_{max}} d\cos\theta'\,\left( R_1(\theta',q')+R_2(\theta',q') \right) + R_3(\theta',q')
\end{equation}
where 
\begin{subequations}
 \begin{eqnarray}
  R_1(\theta',q') & = & - \imath f_{\theta'}
\left(\left[ \dot{H}^{(f)}, \rho_{{\nu}_{\underline{\alpha}}}(q',\theta')\right] L_{{\nu}_{\underline{\alpha}}}(q') \right.\nonumber\\
 & & \left. \qquad + \left[ \dot{\bar{H}}^{(f)}, \rho_{{\bar{\nu}}_{\underline{\alpha}}}(q',\theta')\right]^* L_{\bar{{\nu}}_{\underline{\alpha}}}(q')\right) \\
 R_2(\theta',q') & = & - \imath f_{\theta'} \left(-\imath\left[ H^{(f)}, \left[ H^{(f)}, \rho_{{\nu}_{\underline{\alpha}}}(q',\theta')\right] \right] L_{{\nu}_{\underline{\alpha}}}(q')\right. \nonumber\\
 & & \left. \qquad+\imath\left[ \bar{H}^{(f)}, \left[ \bar{H}^{(f)}, \rho_{{\bar{\nu}}_{\underline{\alpha}}}(q',\theta')\right] \right]^* L_{{\bar{\nu}}_{\underline{\alpha}}}(q')\right) \nonumber\\
\\
 R_3(\theta',q') & = & \imath S(x)f_{\theta_{max}} \,\left(\left[ H^{(f)}, \rho_{{\nu}_{\underline{\alpha}}}(q',\theta')\right] L_{{\nu}_{\underline{\alpha}}}(q') \right. \nonumber\\
 & & \left.\qquad+ \left[ \bar{H}^{(f)}, \rho_{{\bar{\nu}}_{\underline{\alpha}}}(q',\theta')\right]^* L_{\bar{{\nu}}_{\underline{\alpha}}}(q')\right)
 \end{eqnarray}
\end{subequations}
with
\begin{equation}
 f_{\theta'} = \left(1-\cos\theta \cos\theta'\right) .
\end{equation}

For the other two one obtains
\begin{eqnarray}
\dot{S}(x)&=& R^2_\nu \left[ -\dfrac{3}{x^4 g(x)} - \dfrac{R^2_\nu}{x^6 g(x)^{3}} \right]
\end{eqnarray} 
with
\begin{equation}
 g(x) = \sqrt{1-\left(R_\nu/x\right)^2}
\end{equation}
and

\begin{equation}
 \dot{T}(q') = T_1(q') + T_2(q')
\end{equation}
with 
\begin{subequations}
 \begin{eqnarray}
T_1(q')&=& -\cos\theta\,S(x)\,\left[\rho_{\nu_{\underline{\alpha}}} (q',\theta_{max}) L_{\nu_{\underline{\alpha}}}(q') \right. \nonumber\\
&&\left.\qquad- \rho_{\bar{{\nu}}_{\underline{\alpha}}}^*(q',\theta_{max}) L_{\bar{{\nu}}_{\underline{\alpha}}}(q')\right] \\
T_2(q')&=&-\imath f_{\theta_{max}} \left[\left[ H^{(f)}, \rho_{{\nu}_{\underline{\alpha}}}(q',\theta_{max})\right] L_{{\nu}_{\underline{\alpha}}}(q') \right. \nonumber\\
&&\left.\qquad+ \left[ \bar{H}^{(f)}, \rho_{{\bar{\nu}}_{\underline{\alpha}}}(q',\theta_{max})\right]^* L_{\bar{{\nu}}_{\underline{\alpha}}}(q')\right] .
\end{eqnarray}
\end{subequations}

\end{document}